# Pressure-induced structure phase transitions and superconductivity in dual topological insulator BiTe

Shihao Zhu[1#], Bangshuai Zhu[1#], Cuiying Pei[1], Qi Wang[1,2], Jing Chen[3], Qinghua Zhang[3], Tianping Ying[3], Lin Gu[4], Yi Zhao[1], Changhua Li[1], Weizheng Cao[1], Mingxin Zhang[1], Lili Zhang[5], Jian Sun[6], Yulin Chen[1,2,7], Juefei Wu[1*], Yanpeng Qi[1,2,8*]

1. School of Physical Science and Technology, ShanghaiTech University, Shanghai 201210, China
2. ShanghaiTech Laboratory for Topological Physics, ShanghaiTech University, Shanghai 201210, China
3. Institute of Physics and University of Chinese Academy of Sciences, Chinese Academy of Sciences, Beijing 100190, China
4. Beijing National Center for Electron Microscopy and Laboratory of Advanced Materials, Department of Materials Science and Engineering, Tsinghua University, Beijing 100084, China
5. Shanghai Synchrotron Radiation Facility, Shanghai Advanced Research Institute, Chinese Academy of Sciences, Shanghai 201203, China
6. National Laboratory of Solid State Microstructures, School of Physics and Collaborative Innovation Center of Advanced Microstructures, Nanjing University, Nanjing 210093, China
7. Department of Physics, Clarendon Laboratory, University of Oxford, Parks Road, Oxford OX1 3PU, UK
8. Shanghai Key Laboratory of High-resolution Electron Microscopy, ShanghaiTech University, Shanghai 201210, China

[#] These authors contributed to this work equally.
[*] Correspondence should be addressed to Y.Q. (qiyp@shanghaitech.edu.cn) or J.W. (wujf@shanghaitech.edu.cn)

**The $(Bi_2)_m(Bi_2Te_3)_n$ homologous series possess natural multilayer heterostructure with intriguing physical properties at ambient pressure. Herein, we report the pressure-dependent evolution of the structure and electrical transport of the dual topological insulator BiTe, a member of the $(Bi_2)_m(Bi_2Te_3)_n$ series. With applied pressure, BiTe exhibits several different crystal structures and distinct superconducting states, which is remarkably similar to other members of the $(Bi_2)_m(Bi_2Te_3)_n$ series. Our results provide a systematic phase diagram for the pressure-induced superconductivity in BiTe, contributing to the highly interesting physics in this $(Bi_2)_m(Bi_2Te_3)_n$ series.**

# 1. Introduction

The successful synthesizing of natural van der Waals (vdW) topological $(MnBi_2Te_4)_m(Bi_2Te_3)_n$ has evoked the attention on studying natural heterostructures[1-4]. They possess intrinsic homogeneous interfaces and tunable properties by stacking layer, such as magnetism and topology[5-8]. Such materials provide a platform for studying Majorana fermions, topological superconductivity and quantum computation[9-12], avoiding the complex process such as metal-organic chemical vapor deposition (MOCVD) and molecular beam epitaxy (MBE) in synthesizing artificial topological heterostructures.

BiTe is a member of the natural multilayer heterostructure in the $(Bi_2)_m(Bi_2Te_3)_n$ ($m$ = 1, $n$ = 2) homologous series[13, 14]. Two $Bi_2Te_3$ quintuple layers (QLs) sandwich a zigzag Bi-Bi bilayer *via* weak vdW interactions, and the building block $Bi_2Te_3$ is a well-known topological insulator (TI)[15-17]. Recently, Eschbach et. al preformed angle-resolved photoemission spectroscopy (ARPES) measurements together with density functional theory (DFT) calculations and demonstrated that BiTe is a dual three-dimensional (3D) topological insulator where a weak topological insulator phase and topological crystalline insulator phase appear simultaneously[18].

In addition, the tunability of natural heterostructures is owning to the interlayer and intralayer interactions, novel properties may occur under external pressure[19-24]. For instance, we discovered pressure-induced superconductivity in natural multilayer heterostructure $(MnSb_2Te_4)_m(Sb_2Te_3)_n$ ($m$ = 1, $n$ = 1)[25] and $(PbSe)_5(Bi_2Se_3)_{3m}$ ($m$ = 2)[26]. Relatively, the high-pressure properties of BiTe are less explored. Since several members of the $(Bi_2)_m(Bi_2Te_3)_n$ (e.g. Bi[27], $Bi_2Te_3$[28], $Bi_4Te_3$[29]) homologous series exhibit distinct superconducting states upon compression, it is intriguing to investigate the structural and electronic properties of BiTe under high pressure.

In this work, we systematically investigate the effect of pressure on the electrical transport and structural properties of natural heterostructure BiTe combining high-pressure *in situ* X-ray diffraction, Raman spectroscopy, electrical transport measurements and theoretical calculations. Under high pressure, BiTe undergoes multiple structural phase transitions: from ambient $P\bar{3}m1$ phase transforms to the high pressure $P2_1/m$ phase around 5 GPa, subsequently, into $Pm\bar{3}m$ phase around 13 GPa. Accompanying these structural phase transitions, two distinct superconducting states are observed with a maximum $T_c$ = 8.4 K at around 11.9 GPa. The theoretical calculations demonstrate that the pressure dependence of the electron phonon coupling (EPC) coefficient $\lambda$ agrees well with the variation of $T_c$. We present a rich phase diagram of BiTe and relate it with the sequential structural transitions.

## 2. Experimental section and Calculating methods

High-quality crystalline sample of BiTe was synthesized by stoichiometric mixing of ultra-pure constituent elements Bi and Te in quartz tube[30]. Subsequently the quartz tube was sealed at ~$10^{-6}$ torr vacuum. After sealing, the quartz tube was slowly heated to 723 K over 12 hours to minimize Te evaporation and then to 1123 K over 5 hours. The tube was then soaked for 6 hours at 1123 K and cooled to 893 K followed by annealing for 6 hours and finally brought to room temperature over a time of 8 hours.

The crystalline phase of BiTe was characterized by high-resolution scanning transmission electron microscope (STEM) and powder X-ray diffraction (XRD). After the single crystal was fully grounded in acetone, the small fragments suspended in acetone were dripped onto a TEM microgrid. The atomic positions of BiTe were characterized using an ARM-200CF (JEOL, Tokyo, Japan) transmission electron microscope operated at 200 kV. The XRD measurements were performed on a Bruker AXS D8 Advance powder crystal X-ray diffractometer with Cu $K_{\alpha 1}$ (wavelength $\lambda$ = 1.54178 Å) at room temperature. The atomic ratio was characterized by the energy dispersive X-ray spectrometry (EDXS).

A piece of BiTe crystal was loaded into diamond anvil cell (DAC) for high-pressure electrical transport measurements using van der Pauw method[31-34]. Four platinum sheet electrodes were touched to the sample for resistance measurements. High-pressure *in situ* XRD experiments were performed at BL15U1 beamline of Shanghai Synchrotron

Radiation Facility (wavelength $\lambda$ = 0.6199 Å). A symmetric DAC with 400 μm culets and T301 gasket was used. The diffraction images were integrated azimuthally with the Fit2D software to yield conventional intensity vs $2\theta$ diffraction diagrams and then analyzed with Rietveld method by using program General Structure Analysis System (GSAS) and the graphical user interface EXPGUI. High-pressure *in situ* Raman spectroscopy investigations were carried out on a Raman spectrometer (Renishaw in Via, U.K.) with a laser excitation wavelength of 532 nm as well as a low-wavenumber filter. Mineral oil was used as pressure transmitting medium. Pressure in the experiments was calibrated by the ruby luminescence method[35].

We carried out the machine learning and graph theory accelerated crystal structure search method (Magus)[36] to identify the enthalpy lowest structures of BiTe in the pressure range of 0-30 GPa. Our first-principles calculations were carried out by utilizing the Vienna *ab initio* Simulation Package (VASP)[37] within the framework of DFT[38, 39]. The Perdew–Burke–Erzernhof (PBE) functional based on the generalized gradient approximation (GGA)[40, 41] was chosen to describe the exchange–correlation interaction, and the projector augmented wave (PAW)[42] method was adopted with the energy cutoff of plane-wave basis set at 450 eV. The convergence criterions for geometry optimization and atomic relaxation were $3\times 10^{-3}$ eV/Å and $1 \times 10^{-6}$ eV per atom for force and energy, respectively. A Monkhorst-Pack[43] *k*-point grids with a reciprocal spacing $2\pi \times 0.03$ Å$^{-1}$ in the Brillouin zone was selected. The vdW effects were taken into account using the Becke-Johnson damping DFT-D3 method[44]. Phonon spectrum calculations were performed by utilizing the supercell finite displacement method implemented in the PHONOPY package[45]. $3 \times 3 \times 3$ and $4 \times 4 \times 4$ supercells were applied to the *P*2$_1$/*m* and *Pm*$\overline{3}$*m* phases, respectively. The EPC coefficient $\lambda$ were calculated by the QUANTUM-ESPRESSO (QE) package[46] using density-functional perturbation theory[47]. We selected the ultrasoft pseudopotential with a kinetic energy cutoff of 70 Ry. We used the Eliashberg equationand Allen-Dynes modified McMillan equation[48-50] to estimate the superconducting transition temperature $T_c$.

$$T_c = \frac{\omega_{\log}}{1.2}\exp\left(\frac{-1.04(1+\lambda)}{\lambda-\mu^*(1+0.62\lambda)}\right) \quad (1)$$

where $\omega_{log}$ is the logarithmic average frequency, and $\mu^*$ is the Coulomb pseudopotential for which we used the value 0.10.

## 3. Results and discussion

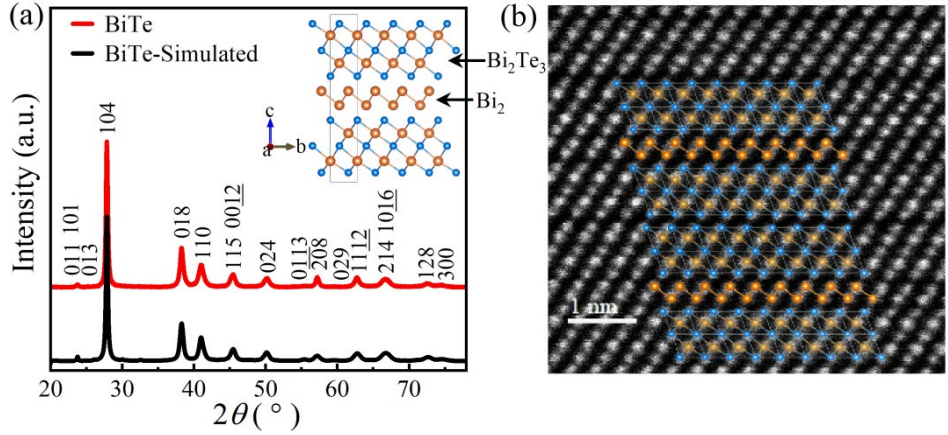

FIG.1 (a) Powder X-ray diffraction (XRD) pattern of BiTe at ambient pressure. Inset: crystal structure of BiTe. (b) Angular dark-field (ADF) image of BiTe. Aberration-corrected STEM image of BiTe with well-organized atomic position along [001] (Bi atom: orange; Te atom: blue).

At ambient pressure, BiTe crystallizes in a hexagonal structure with space group $P\bar{3}m1$ (No.164). Prior to high-pressure measurements, we first checked sample quality by powder XRD, EDXS and STEM. As shown in Fig. 1(a), all the reflections in the powder XRD patterns can be indexed based on pure BiTe phase. The extracted lattice parameters are $a$ = 4.3924 Å and $c$ = 23.8405 Å, in agreement with previous reports[30]. From the EDXS data in the inset of Fig. S1, the chemical composition Bi : Te = 1 : 0.94, indicating slight Te vacancies in the as-grown sample. An annular dark-field STEM (ADF-STEM) image of BiTe along [001] is shown in Fig. 1(b). It is clear to see that a zigzag chain of Bi-Bi bilayer is sandwiched between two layers of $Bi_2Te_3$. An alternate stack of $Bi_2Te_3$ and Bi-Bi along $c$-axis builds a multilayer heterostructure in a hexagonal unit cell. The above characterizations indicate a high quality of our samples.

Next, we carried out transport measurements at ambient pressure. As shown in Fig. S1, BiTe shows typical metallic behavior from 1.8 K to 300 K. Theoretical calculations indicated that BiTe is a dual 3D topological insulator with a narrow band gap at ambient pressure. Our chemical compositional analysis illustrates that $BiTe_{0.94}$ has 6% Te deficiency. Such defects could lead to highly electron doping, shifting Fermi level into conduction band and causing metallic conduction[28, 51]. This typical metallic behavior was also observed in other topological insulators $Bi_2Te_3$[28] and BiSe[52].

The high-pressure approach has been widely employed in recent studies of topological materials and has led to many interesting results[53-56]. Hence, we investigated the effects of high-pressure on electrical transport properties of BiTe crystals. Figs. 2 (a) and (b) show the temperature-dependent resistivity of BiTe at various pressures. At 1.0 GPa, the resistivity of BiTe in the whole temperature range shows a metallic nature, which is similar to that at ambient pressure. Increasing pressure induces a continuous suppression on the overall magnitude of resistivity, and a superconducting transition emerges at 4.1 GPa combining with zero resistivity. The superconducting transition temperature $T_c$ (90% drop of the normal state resistivity) reaches 2.6 K and the superconducting transition width $\Delta T_c$ (10%-90% of the normal state resistivity at $T_c$) is ~ 0.4 K. With further increase of pressure, $T_c$ increases rapidly, however, $\Delta T_c$ becomes broadened. A maximum $T_c$ of 8.4 K is attained at around 11.9 GPa. Beyond this pressure, $T_c$ decreases slowly, but $\Delta T_c$ becomes sharp again.

To confirm the superconducting transition, we applied external magnetic fields to sample under 11.9 GPa and 29.1 GPa, respectively. As shown in Figs. 2 (c) and (d), $T_c$ is suppressed with the enhancement of magnetic fields and the superconductivity extinguishes under the magnetic field $\mu_0 H$ = 3 T and 1.6 T, respectively. We used Ginzburg-Landau formula to fit the data [Figs. 2 (e) and (f)]. The estimation of $\mu_0 H_{c2}(0)$ is ~ 2.5 T and 1.5 T for 11.9 GPa and 29.1 GPa, respectively. It should be noted that the $\mu_0 H_{c2}(0)$ obtained here is lower than its corresponding Pauli paramagnetic limit $H_P$ = 1.84$T_c$. According to the relationship $\mu_0 H_{c2} = \Phi_0/(2\pi\xi^2)$, where $\Phi_0 = 2.07 \times 10^{-15}$ Wb is the flux quantum, the Ginzburg-Landau coherence length $\xi_{GL}(0)$ is 11.5 nm at 11.9 GPa and 14.8 nm at 29.1 GPa, respectively.

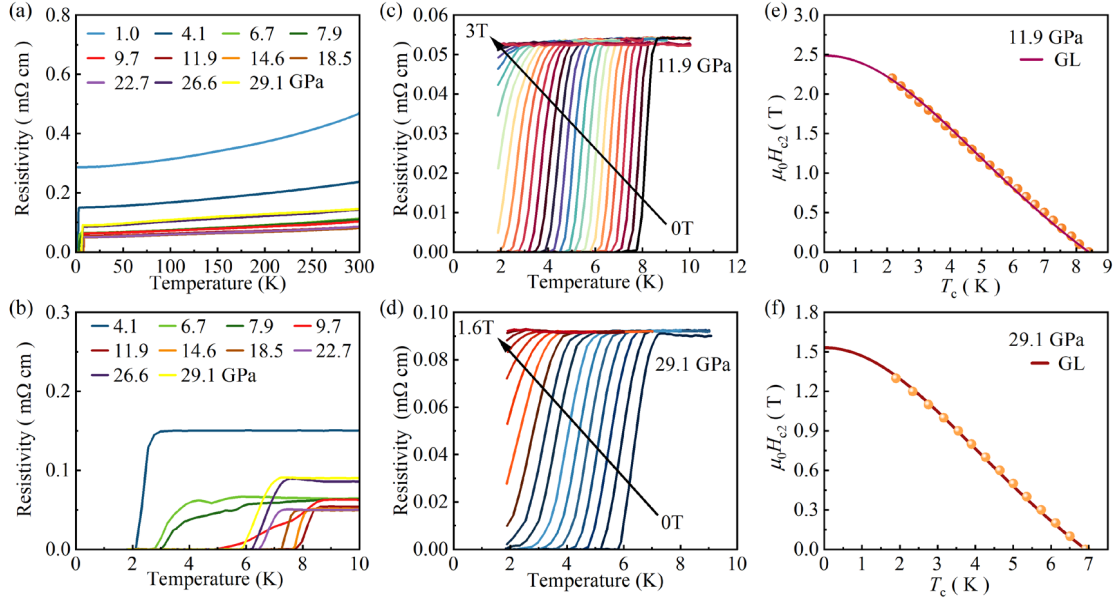

FIG.2 Transport properties of BiTe as a function of pressure. (a) Electrical resistivity as a function of temperature within 29.1 GPa in run III. (b) Temperature-dependent resistivity of BiTe in the vicinity of superconductivity. Temperature dependence of resistivity under different magnetic fields for BiTe at 11.9 GPa (c) and 29.1 GPa (d), respectively. The results of Ginzburg-Landau fitting are shown in (e) and (f).

Since $T_c$ displays a non-monotonic evolution with increasing pressure and $\Delta T_c$ drastically changes at around 6.7 ~ 9.7 GPa, we assume that there is alternation on the nature of the superconducting state under high pressure. Given that the $(Bi_2)_m(Bi_2Te_3)_n$ homologous series usually exhibits several different crystal structures under high pressure, we first performed crystal structure searching based on first-principles calculations at 10 GPa and 20 GPa, respectively. Structural evolutions were implemented for 25 generations with 30 structures per generation, and the ambient $P\bar{3}m1$ phase was treated as seeds during the structure searches. We found two stable candidates under high pressure: the monoclinic $P2_1/m$ phase and the cubic $Pm\bar{3}m$ phase, as shown in Fig. 3(a). To investigate the thermodynamic stabilities of the newly identified structures, we calculated the relative enthalpy difference with respect to the ambient $P\bar{3}m1$ phase in the pressure range of 0-30 GPa, as plotted in Fig. 3(b). The enthalpy calculations suggest that BiTe undergoes two structural phase transitions under high pressure. The ambient $P\bar{3}m1$ phase transforms to the predicted $P2_1/m$ phase at around 5.7 GPa. Then, as depicted in the inset of Fig. 3(b), the predicted $Pm\bar{3}m$ phase is more energetically stable than $P2_1/m$ phase above 7.2 GPa. The relative enthalpy curves of $P2_1/m$ and $Pm\bar{3}m$ phases almost overlap with each other above 15 GPa. In particular, the enthalpy difference between $P2_1/m$ and $Pm\bar{3}m$ phases is 0.34 meV/atom at 15 GPa, which is beyond the convergence limitation of our calculations. We look into the structures and find that $P2_1/m$ phase transforms to $Pm\bar{3}m$ phase after structural optimization above 15 GPa, which further suggests the stability of $Pm\bar{3}m$ phase under high pressure.

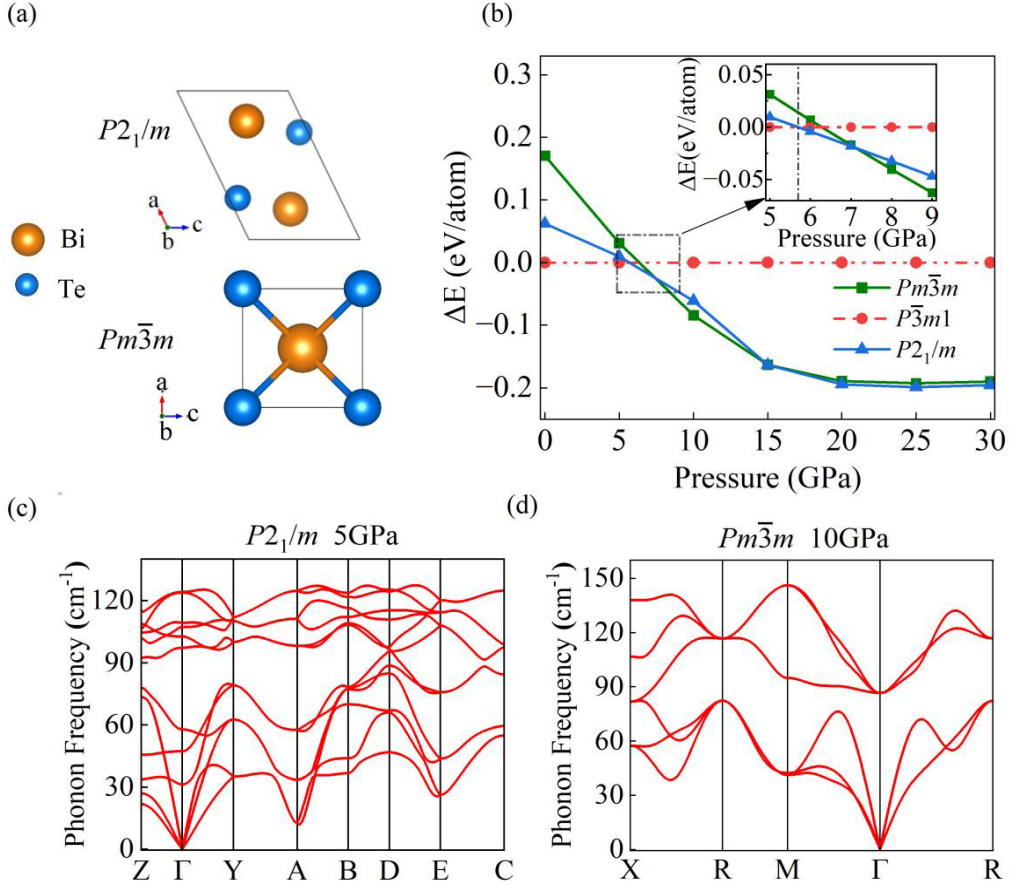

FIG.3 (a) The predicted $P2_1/m$ and the $Pm\bar{3}m$ structures. (b) The calculated enthalpy difference as a function of pressure for $P\bar{3}m1$, $P2_1/m$ and $Pm\bar{3}m$. The values are relative to $P\bar{3}m1$. (c) The calculated phonon spectra of the $P2_1/m$ phase at 5 GPa and (d) the $Pm\bar{3}m$ phase at 10 GPa.

In addition, we calculated the phonon dispersions of the predicted $P2_1/m$ phase and $Pm\bar{3}m$ phase. There are no imaginary frequencies along the whole Brillouin path for $P2_1/m$ phase from 5 GPa [Fig. 3(c)] to 10 GPa [Fig. S4(f)], and $Pm\bar{3}m$ phase from 10 GPa [Fig. 3(d)] to 30 GPa [Fig. S4(c)], illustrating their dynamical stability under high pressure. Nevertheless, both phonon spectra of $P2_1/m$ and $Pm\bar{3}m$ phases have imaginary frequencies at ambient pressure [Fig. S4(d) and Fig. S4(a)], indicating that both the high-pressure phases may not persist after pressure releasing.

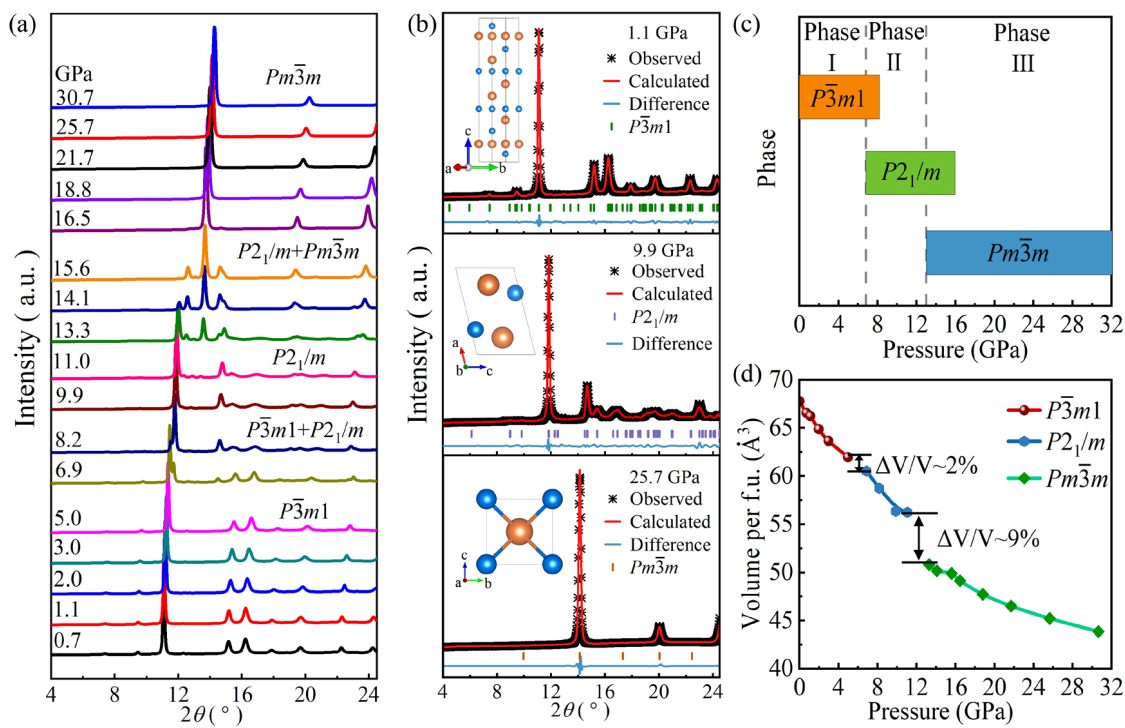

FIG.4 (a) High pressure XRD patterns of BiTe up to 30.7 GPa at room temperature. The X-ray diffraction wavelength λ is 0.6199 Å. (b) Rietveld refinement of XRD pattern at 1.1 GPa, 9.9 GPa and 25.7 GPa. The red solid line and black stars represent the calculated and experimental data, respectively, and the azure lines are the residual intensities. The vertical bars are the diffraction peak positions. (c) The phase transition sequence upon pressure increase. (d) Pressure dependence of volume per formula unit in different phases.

To confirm the predicted structure transitions, we performed high-pressure *in situ* powder XRD measurements up to 30 GPa at room temperature, as displayed in Fig. 4(a). At low pressure region, all the diffraction peaks can be indexed well to the ambient hexagonal $P\bar{3}m1$ structure (phase I). A representative refinement at 1.1 GPa is displayed in Fig. 4(b). The ambient phase is robust and can remain till 5.0 GPa. At 6.9 GPa, the diffraction pattern suddenly changes, signifying the formation of phase II. This phase is only stable in a narrow pressure range and coexists with the phase I or the phase III upon compression. When the pressure reaches 13.3 GPa, new Bragg peaks emerge, indicating that phase II is transformed into phase III. Above 16.5 GPa, only phase III exists and no further transitions are observed up to 30.7 GPa. The observed phase transitions agree with our theoretical prediction. Furthermore, we have refined the observed XRD data at 9.9 GPa and 25.7 GPa using the predicted $P2_1/m$ and $Pm\bar{3}m$ phases, respectively. The uses of the predicted phases gave excellent Rietveld fittings [Fig. 4(b)], hence leading to the unambiguous determination of phase II and phase III as the predicted $P2_1/m$ and $Pm\bar{3}m$ phases, respectively. Fig. 4(d) shows the experimental volume data as a function of pressures for phases I, II, and III. It is obvious that the two phase transitions can be characterized by first order accompanying 2% and 9% volume drops at the transitions, respectively.

The pressure-induced structure evolution of BiTe was also confirmed by *in situ* Raman spectroscopy measurements. As shown in Fig. S3, the signal of $A_{1g}^1$ disappears around 5.4 GPa, and new vibration mode (150 cm$^{-1}$) appears at 9.6 GPa, indicating the structural transitions under high pressure. Therefore, the results of synchrotron XRD and Raman spectroscopy measurements provide additional evidence of the pressure-induced structural transitions in BiTe, which agrees with our theoretical structure predictions. The aforementioned results demonstrate that application of pressure dramatically alters both crystal and electronic structures of BiTe, similar to other members of $(Bi_2)_m(Bi_2Te_3)_n$ homologous series, like Bi, $Bi_2Te_3$ and $Bi_4Te_3$. The details are summarized in Table I.

We performed several independent high-pressure transport measurements on BiTe crystals and provided consistent and reproducible results, confirming the intrinsic superconductivity of BiTe under high pressure (Fig. S2). Based on the aforementioned results, we can establish a $T_c$-$P$ phase diagram for BiTe shown in Fig. 5. Superconductivity is observed at around 5 GPa. $T_c$ increases sharply up to around 12 GPa and then decreases slowly. A dome-like evolution is displayed with a maximum $T_c$ of 8.4 K at 11.9 GPa. The high-pressure *in situ* synchrotron XRD and Raman spectroscopy reveal the evidence of structural transitions around 5.0 GPa and 12.0 GPa respectively, which is in line with the theoretical predictions that the ambient phase I transforms to the high-pressure phase II and phase III. Occurrence of superconductivity at around 4-6 GPa is associated with the pressure-induced monoclinic phase. Moreover, we assume that the broadening of $\Delta T_c$ from 6.7 to 9.7 GPa originates from the mixing of phase II and phase III.

**TABLE I**. Summary of phases observed in $(Bi_2)_m(Bi_2Te_3)_n$ homologous series under high pressure. SC = superconductivity observed.

|  | Bi[27] | BiTe | $Bi_2Te_3$[28, 57] | $Bi_4Te_3$[29] |
|---|---|---|---|---|
| Phase I | $R\bar{3}m$ <br> 0-2.5 GPa <br> No SC | $P\bar{3}m1$ <br> 0-7 GPa <br> No SC | $R\bar{3}m$ <br> 0-8 GPa <br> SC (3.2-8 GPa) | $R\bar{3}m$ <br> 0-6.4 GPa <br> No SC |
| Phase II | $C2/m$ <br> 2.5-2.6 GPa <br> SC | $P2_1/m$ <br> 6-15 GPa <br> SC | $C2/m$ <br> 8-12 GPa <br> SC | $C2/m$ <br> 6.4-11.5 GPa <br> SC |
| Phase III | $I4/mcm$ <br> 2.7-7.7 GPa <br> SC | $Pm\bar{3}m$ <br> >13 GPa <br> SC | $C2/c$ <br> 12-14 GPa <br> SC | <br> 11.5-16 GPa <br>  |
| Phase IV | $Im\bar{3}m$ <br> >7.7 GPa <br> SC |  | $I\bar{3}m$ <br> >14 GPa <br> SC | $Im\bar{3}m$ <br> >16 GPa <br> SC |

In order to understand the physical properties of BiTe, we calculated the electronic structures under different pressures, as shown in Fig. S5 and Fig. S6. At lower pressure region, the energy gap of ambient phase I increases to 0.2 eV at 4 GPa, and then decreases at 5 GPa. Different from the ambient phase, both phase II and phase III are metallic and exhibit finite density of states (DOS) at the Fermi energy ($E_F$). Electrons of Bi atoms make main contribution to DOS in both phase II and phase III around $E_F$. It is worth noting that there is a peak around 0.5 eV in phase III which could be related to the Dirac-like crossing along the Brillouin path X-R, as plotted in the Figs. S6 (e) and (f). Moreover, we calculated the EPC coefficient $\lambda$ and $T_c$ value of phase II and phase III under high pressure, as depicted in Fig. S7. Both phase II and phase III are superconducting and the $T_c$ value has a dome-like behavior with the maximum of 5.6 K around 10 GPa. The overall trend of the calculation is in line with the $T_c$ measurements under high pressure.

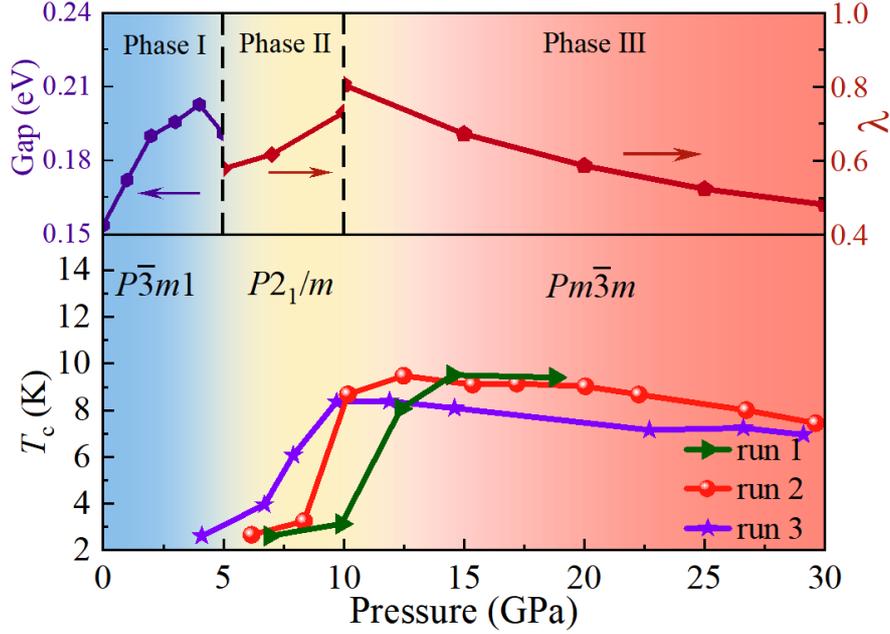

FIG.5 Electronic phase diagram of BiTe. The colored areas represent different phases, shaded areas indicate the coexistence of two phases. The upper panel shows the pressure dependence of the calculated band gaps of $P\bar{3}m1$ phase, and the electron phonon coupling coefficient $\lambda$ for $P2_1/m$ and $Pm\bar{3}m$ phases. The lower panel shows the superconducting $T_c$ as a function of pressure.

## 4. Conclusion

In summary, we have performed experimental and theoretical studies on the electrical transport properties and crystal structures of BiTe under high pressure. Application of pressure effectively tunes both crystal and electronic structures of BiTe. Two pressure-induced superconducting states are observed upon compression, which are related to structural transitions as evidenced by both the synchrotron XRD and Raman spectroscopy measurements. Given that all the members of the $(Bi_2)_m(Bi_2Te_3)_n$ (e.g. Bi, $Bi_2Te_3$, $Bi_4Te_3$) homologous series exhibit several different crystal structures and distinct superconducting states, the $(Bi_2)_m(Bi_2Te_3)_n$ family provides an extensive series that help to illuminate the subtle differences that create topologically insulating or superconducting states.

**Acknowledgements**

This work was supported by the National Natural Science Foundation of China (Grant Nos. 52272265, U1932217, 11974246, 12004252), the National Key R&D Program of China (Grant No. 2023YFA1607400, 2018YFA0704300), and Shanghai Science and Technology Plan (Grant No. 21DZ2260400). The authors thank the Analytical Instrumentation Center (# SPST-AIC10112914), SPST, ShanghaiTech University and the Analysis and Testing Center at Beijing Institute of Technology for assistance in facility support. The authors thank the staffs from BL15U1 at Shanghai Synchrotron Radiation Facility for assistance during data collection.